%Paper: hep-th/9308055
%From: TEMPLE@Vax2.Concordia.CA
%Date: Wed, 11 Aug 1993 14:18:08 -0500 (EST)
%Date (revised): Thu, 12 Aug 1993 09:51:39 -0500 (EST)

\magnification\magstep1
\def\curv{{\rm\Omega}}
\noindent
\centerline{\bf Instantons in topological field theories}
\vskip 0.3in
\centerline{M. Temple-Raston}
\vskip 0.2in
\centerline{Department of Mathematics and Statistics,}
\vskip 0.1in
\centerline{Concordia University, 1455 de Maisonneuve Blvd. W.,}
\vskip 0.1in
\centerline{Montr\'eal, Qu\'ebec, Canada H3G 1M8}
\vskip 0.5in
\noindent
{\bf Abstract}
\vskip 0.1in
\noindent
On an oriented, compact, connected, real four-dimensional manifold,
$M$, we introduce a topological Lagrangian gauge field theory
with a Bogomol'nyi structure that leads to non-singular,
finite-Action, stable solutions to the variational field equations.
These soliton-like solutions are analogous to the instanton in
Yang-Mills theory.  Unlike Yang-Mills instantons, however, `topological'
instantons are independent of any underlying metric structure, and, in
particular, they are independent of the metric signature.  We show that
when the topology of the underlying manifold, $M$, is equipped with a
complex K\"ahler structure, and $M$ is interpreted as space-time, then
the moduli space of topological instantons---the space of
motions---is a finite-dimensional, smooth, Hausdorff manifold with
a natural symplectic structure.  We identify space-time topologies
which lead to the physical stability of topological instanton
field configurations compatible with the additional geometric
structures.  The spaces of motion for $U(1)$ topological
instantons over either minimal elliptic or algebraic complex
space-times with irregularity $q=2$ are examined.
\vskip 0.5in
\openup 2\jot
\noindent
{\bf 1. Introduction}
\medskip
\noindent
Witten gave the first examples of topological field theories
(TFTs) [W1,W2], and Horowitz proposed a general class of TFTs
which included some of Witten's examples as special cases [H].
To date, most of the interest in TFTs has been in the quantum
field theory, where deep connections have been forged with the new
smooth-invariants of low dimensional manifolds (e.g., Donaldson
invariants) [W1,D1].  In this paper we present and study a class
of topological gauge field theories which have a `minimising'
Bogomol'nyi structure.  The TFTs we introduce below are classical
covariant Lagrangian Action functionals with a Bogomol'nyi
structure over oriented, compact, connected real four-manifolds.
The functionals do not depend on the metric structure of the
underlying space-time (classical `general' covariance).
The Bogomol'nyi structure leads to first-order field equations,
the Bogomol'nyi equations, with solutions that automatically
satisfy the variational field equations, and, that are stable
under perturbation.  The Bogomol'nyi equations studied in this
paper are metricless versions of the (anti-)self-duality
equations of Yang-Mills gauge theory.
The TFTs introduced below are Horowitz's TFTs
generalised to strengthen the Bogomol'nyi structure for all
ranks of vector bundle.
In this paper we shall not discuss the quantization of the TFTs,
because we believe that the classical Bogomol'nyi soliton
is sufficiently interesting to be examined on its own.
Our view comes from recent work which suggests that
classical Bogomol'nyi solitons can have remarkable
`pseudo-quantum' dynamics [TRA].

\smallskip
We now summarise the sections of this paper.
In the next section we introduce classical generally covariant
Lagrangian field theories over an oriented, compact, connected real
four-manifold with a natural Bogomol'nyi structure.  Because
of the similarity of the topological Bogomol'nyi equations to
the self-duality equations in Yang-Mills theory, the non-singular,
finite-Action, stable solutions to the topological Bogomol'nyi
equations are called `topological instantons'.
In section three, we study the line bundle case and introduce further
geometrical structure onto the theory: we assume that the underlying
compact, connected base-manifold is equipped with
a K\"ahler structure.  We obtain a natural correspondence between
topological instantons and Einstein-Hermitian vector bundles.
This leads to a finite-dimensional, smooth, Hausdorff manifold
which we can interpret as the space of motions later, in section five.
In section four the appropriate generalisation to vector bundles
is presented.  In sections three and four we make use of recent
results achieved by mathematicians.
The differential geometric notion of stability
introduced by Kobayashi [Ko1], and the work of Kim [Kim] that
studied Einstein-Hermitian moduli spaces, are particularly useful.
Section five views the underlying K\"ahler manifold
as space-time, so that the moduli space of Einstein-Hermitian
topological instantons becomes the space of motions.  We argue
that the dynamics of Bogmol'nyi solitons in TFTs may not be
trivial, and that diagonal abelian Einstein-Hermitian topological
instantons may correspond to either photons, massless neutrinos,
or composite lumps of photons or neutrinos.  Section six
provides a conclusion.

\bigskip
\noindent
{\bf 2. Topological instantons}

\medskip
Let $\pi\colon P\to M$ be a principal $G$-bundle over an oriented,
compact, connected four manifold $M$, and denote by $E=P\times_G{\cal
G}$, the vector bundle associated to $\pi\colon P\to M$ by the adjoint
representation of $G$ on the Lie algebra, ${\cal G}$.  Denote by
${\cal A}(P)$ the space of $L^2_1$ connections on $P$, and
consider {\it two} connections, $A,B\in{\cal A}(P)$.
$A$ and $B$ induce exterior covariant derivatives $D^A$ and $D^B$ on
the associated vector bundle $E$.  To define the curvatures of $D^A$
and $D^B$, let $s$ be a local frame field of $E$
over an open set $U\subset M$.
The curvatures $H^A$ and $K^B$ are given by $D^AD^As=sH^A$
and $D^BD^Bs=sK^B$ and can be interpreted as 2-forms on $M$ taking values
in $E$, i.e., $H^A,K^B\in\Omega^2(M,E)$. Finally, we assume that there is
an invariant positive-definite inner product $<\ ,\ >$ on $E_x$ which
varies continuously with $x\in M$.

\smallskip
The topological field theories we study in this paper differ from
those introduced by Horowitz in [H] for theories of gravity.
The modifications in this paper permit a different geometrical
formulation of the problem.  The Lagrangian theories of interest
to us are given by the functional
$$\eqalign {{\cal L}_{\pm}(A,B)=\int_M\,<(H^A\otimes I_E)&\wedge
(I_E\otimes K^B)>\cr
&\pm{1\over 2}\int_M\,<(I_E\otimes K^B)\wedge (I_E\otimes K^B)>,}\eqno{(1)}$$
defined on the product space ${\cal A}(P)\times{\cal A}(P)$.  Interpreting
$H^A$ and $K^B$ as curvatures in the Lagrangian Action requires that
the real dimension of $M$ be four (in [H], $K^B$ is viewed as
an axion potential).  $I_E$ denotes the identity
transformation on the adjoint bundle, $E$.  The brackets $<\ >$ in (1)
remind us that a choice of adjoint-invariant, real-valued inner product
on the adjoint bundle, $E$, has been made.  The variational equations for
(1) are
$$D^AK^B=0, \ \ \ \ D^BH^A=0,\eqno{(2)}$$
where we have made use of the Bianchi identity $D^BK^B=0$.
The set of solutions is clearly neither empty nor entirely trivial.

\smallskip
The physical stability of a class of nontrivial, nonsingular, finite-Action
solutions to the variational equations (3) can be argued when the
Lagrangian is rewritten as
$$\eqalign {2{\cal L}_\pm=\pm\int_M<(H^A\otimes I_E\ \pm\ I_E\otimes K^B)
&\wedge (H^A\otimes I_E\ \pm\ I_E\otimes K^B)>\cr
&\mp\int_M<(H^A\otimes I_E)\wedge (H^A\otimes I_E)>.}\eqno{(3)}$$
The bundle metric defines an adjoint invariant, symmetric, bilinear
real-valued function, $f$, on $E$---that is, a Weil polynomial of
degree two.  When the polynomial is evaluated on the curvature of a
vector bundle $E\to M$, and integrated over a closed $M$, the expression
becomes independent of the connection and represents a topological
invariant for the bundle.  Let $E_A$ and $E_B$ be the vector bundle
$E$ equipped with either the connection $A$ or $B$, respectively.
The first term in the Lagrangian ${\cal L}_-$ in equation (3) can be
interpreted as a topological invariant for the tensor product bundle
$E_A\otimes E_B^*$.  We recall that the curvature of
$E_A\otimes E_B^*$ is given by
$\curv_{E_A\otimes E_B^*}=H^A\otimes I_r-I_r\otimes K^B$.
Therefore the Bogomol'nyi equations,
$$H^A\otimes I_r= I_r\otimes K^B,\eqno{(4)}$$
can be viewed as giving a vanishing curvature condition on the tensor
bundle $E_A\otimes E_B^*$.  Solutions to (4) automatically satisfy the
variational field equations (2).

\smallskip
For line bundles, $L$, solutions to the Bogomol'nyi equations consist
of all pairs of Hermitian connections $(A,B)$ on $L$ for which
$H^A=K^B$.  The Lagrangian in this case reduces to the topological
field theory studied by Baulieu and Singer [BS].  For $r>1$, an
indice computation for the Bogomol'nyi equations (4),
$H_{ab}^A\delta_{cd}=\delta_{ab}K^B_{cd}$, shows that the curvature
forms $H^A$ and $K^B$ are projectively flat.  That is,
$H_{aa}^A=K_{cc}^B=iF$ and $H_{ab}^A=K_{ab}^B=0$ for $a\ne b$, or,
equivalently,
$$H^A=K^B=iF I_r,\eqno{(5)}$$
where $F$ is a real-valued two form on $M$.  For the line bundle,
equation (5) is also valid.  The Bianchi identity for all ranks of
vector bundle imposes a simple condition on $F$, that $dF=0$.  Therefore
$F$ defines a de Rham cohomology class $[F]\in H^2(M,{\bf R})$.
Since $M$ is compact, $H^2(M,{\bf R})$ is of finite dimension.
If $F$ is a curvature on $M$, then the second term in (3) is a
topological invariant of the underlying four-manifold, $M$.  As a
result both terms in ${\cal L}_-$ are topological invariants and
under perturbations the action remains unchanged.
Topologically non-trivial solutions to the Bogomol'nyi
equations will be said to be `physically stable' if $F$ is a
curvature of $M$ and if the solutions have a fixed non-zero action
given by
$${\cal L}_-=-\int_M F\wedge F=-24\pi\ {\rm sgn}(M)\ne 0.\eqno{(6)}$$
As a result ${\cal L}_-$ is proportional to a topological invariant,
the signature of $M$, ${\rm sgn}(M)$.
For ${\cal L}_+$ a similar argument can be made on the tensor bundle
$E_A\otimes E_B$ (instead of on $E_A\otimes E_B^*$),
however here $H^A$ and $K^B$ must be trivially flat.
Flat connections have been well-studied in the literature [W1],
therefore we shall say no more about ${\cal L}_+$.
By analogy with (anti-) self-dual instantons in
Yang-Mills theory and using the physical stability of (6):
\medskip
\noindent
{\bf Definition.}\ \ \ \ {\it A physically stable, non-trivial solution
$(A,B)$ to the Bogomol'nyi equations (5) on the vector bundle $(E,h)$ is
called a topological instanton.}

\medskip
\noindent
Remark.
\smallskip
\noindent
1.  By design, topological instantons are not coupled to the background
space-time curvature.  Consider $r=1$, then a topological instanton,
$A$, has bundle curvature $H^A=iF$, where $F\in H^2(M,{\bf R})$.
If we interpret $F$ as a Maxwell field (the Faraday tensor), then
$dF=0$ is the first half of the Maxwell equations.  The second half of
the Maxwell equations, $\star d\star F=j$,
couples electromagnetism (photons) to the space-time
curvature (gravity), where $j$ is a four-current and the star operator
is defined using the space-time metric structure.  Since a Maxwell field
is also the field of an $r=1$ topological instanton (it is closed),
it is natural to propose the same equation, $\star d\star F=j$,
to couple topological instantons with $r\ge 1$ to the space-time curvature.
The Maxwell equation $\star d\star F=j$ cannot be derived from
a classically generally covariant TFT, and therefore must be introduced
by hand.
Although there is no reason to believe that there is a unique coupling,
it is sufficient for $F$ to be a Maxwell field for $H^A=iFI_E$
($r>1$) to satisfy the non-homogeneous Yang-Mills equations,
$\star D^A\star H^A=J\equiv jI_E$.

\bigskip
\noindent
{\bf 3. Einstein-Hermitian topological instantons on line bundles}

\medskip
In this section we examine the moduli space of $U(1)$ solutions to
the Bogomol'nyi equations (4) when $M$ is a compact, complex
K\"ahler surface.  The moduli space will be interpreted
as the covariant phase space (the space of motions [Sou]),
therefore the moduli space must be topologically
well-behaved (e.g., Hausdorff).  Good topological behaviour can
be achieved by introducing further geometrical structure and by
insisting that the solutions correspond to `stable vector bundles'.
The `stability' of vector bundles, although usually defined
algebro-geometrically, can be phrased completely within
differential geometry.  As we shall see, in differential
geometry stable vector bundles are closely related to
the topological instantons obtained in the previous section.

\smallskip
We set the notation for a general vector bundle, $E\to M$.
The group of bundle automorphisms on $E$ will be denoted by $GL(E)$,
and the subgroup of $GL(E)$ preserving the Hermitian structure, $h$,
by $U(E,h)$.  Let ${\cal D}(E,h)$ be the set of connections $D$ on $E$
preserving $h$, i.e., the set of Hermitian connections.  We equip
$(E,h)$ with a fixed holomorphic structure, $\bar{\partial}_E$.
A holomorphic structure, $\bar{\partial}_E$, and a Hermitian structure,
$h$, together determine a unique compatible connection,
$D=\partial_E+\bar{\partial}_E\in{\cal D}(E,h)$, on
$(E,\bar{\partial}_E,h)$.  Let ${\cal D}^{1,1}(E,h)$ denote the
set of all holomorphic structures on $(E,h)$, or, equivalently, the
subset of Hermitian connections
$D=\partial_E+\bar{\partial}_E\in{\cal D}(E,h)$ which satisfy
$\bar\partial_E^2=0$.  Finally, we recall the
definition of an Einstein-Hermitian vector bundle.  A holomorphic
Hermitian vector bundle $(E,\bar{\partial},h)$ over a complex
Hermitian manifold $(M,g)$ has a weak-Einstein-Hermitian structure if
the unique compatible connection, $D$, has mean curvature, $K(D)$,
which satisfies
$$K(D)=\varphi\,I_E,\eqno{(7)}$$
where $\varphi$ is a real function defined on $M$.  Equation (7) is
called the weak-Einstein condition.  If $\varphi=c$ in (7) with $c$ a
constant, then $D$ is said to satisfy the Einstein condition and
$(E,\bar{\partial}_E,h)$ is an Einstein-Hermitian vector bundle.
Let ${\cal E}(E,h)$ denote the set of Einstein-Hermitian connections,
that is, the connections in ${\cal D}^{1,1}(E,h)\subset {\cal D}(E,h)$
which satisfy equation (7) with $\varphi=c$.
The moduli space of holomorphic structures on $(E,h)$ is given
by ${\cal D}^{1,1}(E,h)/U(E,h)$.  The moduli space of
Einstein-Hermitian structures on $(E,h)$ is given by
${\cal M}_r\equiv{\cal E}(E,h)/U(E,h)$.

\smallskip
Although the notation set above is that for a general vector bundle,
we now specialize to the line bundle case.
Let $(L,{\bar\partial}_L,h)\to(M,g,\Phi)$ be a holomorphic Hermitian
line bundle over a compact, complex K\"ahler surface.  Then, all
connections in ${\cal D}^{1,1}(L,h)$ are Einstein-Hermitian up to a
conformal transformation of the Hermitian structure.  To see this,
we note that every holomorphic Hermitian line bundle
$(L,{\bar\partial}_L,h)$
over $(M,g,\Phi)$ is weak-Einstein-Hermitian for any metric $g$,
since $\Omega=R_{\alpha\bar\beta}\,dz^\alpha\wedge d{\bar z}^\beta$.
Furthermore, there exists a conformal transformation of the Hermitian
structure, $h\to h'=ah$, which makes $(L,{\bar\partial}_L,h')\to(M,g,\Phi)$
an Einstein-Hermitian vector bundle, that is, $K(D)=c$ [Ko2].
The constant $c$ is given by
$$c={2\pi\over V}{\rm deg}(L),$$
where $V={\rm Vol}(M)$ and ${\rm deg}(L)=\int_M\,c_1(L)\wedge\Phi$.
The constant, $c$, is a topological invariant when $M$ is K\"ahler
(when $\Phi$ is closed), and depends only on the cohomology classes of
$\Phi$ and $c_1(L)$.
Therefore all holomorphic Hermitian line bundles $(L,h)\to M$ with
compatible connection are Einstein-Hermitian, up to a
conformal transformation of the Hermitian structure, $h$.
If, in addition, the second cohomology class of $M$ is non-zero,
then from (6) all connections $D\in{\cal D}^{1,1}(L,h)/U(E,h)$
are physically stable, and the converse is trivial: every
Einstein-Hermitian line bundle admits a topological instanton.
We now introduce two types of topological instantons, both compatible
with the K\"ahler and holomorphic structures.
The set of holomorphic K\"ahler topological instantons on
$(L,h)$ is defined by
$${\cal I}^{1,1}(L,h)=\{ (D^A,D^B)\in{\cal D}^{1,1}(L,h)\times
{\cal D}^{1,1}(L,h)\vert (D^A)^2=(D^B)^2=iF,
\ \ \ F\in\Omega^{1,1}(M)\}.$$
Also define the diagonal holomorphic K\"ahler topological instantons
by
$${\cal I}_{\rm dia}^{1,1}(L,h)=\{ (D^A,D^A)\in{\cal D}^{1,1}(L,h)
\times{\cal D}^{1,1}(L,h)\vert (D^A)^2=iF,
\ \ \ F\in\Omega^{1,1}(M)\}.$$
{}From the discussion above, we conclude that
$${\cal I}^{1,1}(L,h)\simeq {\cal E}(L,h)\times{\cal E}(L,h),
\ \ \ \ \
{\cal I}_{\rm dia}^{1,1}(L,h)\simeq {\cal E}(L,h)$$
up to a conformal transformation of the Hermitian structure, $h$.
The isomorphism suggests that we rename ${\cal I}^{1,1}$ and
${\cal I}_{\rm dia}^{1,1}$, Einstein-Hermitian instantons and
diagonal Einstein-Hermitian instantons, respectively.

\smallskip
The moduli space of Einstein-Hermitian structures on Hermitian
vector bundles has been studied extensively by H.J. Kim [Kim,Ko2].
Again restricting to the line bundle, $(L,h)$, Kim proves that
the moduli space ${\cal M}={\cal E}(L,h)/U(L,h)$ is a
nonsingular K\"ahler manifold in a neighbourhood of
$[D]\in{\cal E}(L,h)/U(L,h)$, if $H^2(M,{\rm End}^0(E))=0,$
and when
$${\rm deg}(M)\equiv\int_M c_1(M)\wedge\Phi\ge 0.\eqno{(8)}$$
It is not surprising that the moduli space ${\cal M}$ forms a
well-behaved manifold, because the concept of Einstein-Hermitian
vector bundle was originally introduced by Kobayashi as the
differential geometric counterpart to vector bundle stability [Ko1].
When the inequality (8) is satisfied, the complex dimension
of the moduli space of abelian diagonal Einstein-Hermitian
instantons, if non-empty, can be computed to give
$${\rm dim}_{\bf C}({\cal M})=q,\eqno{(9)}$$
where $q\equiv h^{(0,1)}$ is the irregularity of $M$.  The Hodge
numbers, $h^{(p,q)}$, are defined by
$h^{(p,q)}\equiv{\rm dim}_{\bf C}H^{p,q}(M)$.

\medskip
\noindent
Remarks.
\smallskip
\noindent
1.  Let $M$ be a compact, complex surface with $c_1(M)=0$ and
$b_1(M)=0$ (a K3 surface).
We argue the physical stability of
Einstein-Hermitian instantons from the Noether formula:
$$c_1^2(M)+c_2(M)=12(1-q+p_g).\eqno{(10)}$$
The geometric genus, $p_g$, in equation (10) is defined as the
Hodge number $h^{(0,2)}$.  The irregularity ($q$), the geometric
genus ($p_g$), and the Betti numbers ($b_i\equiv H_i(M,{\bf R})$)
are the homological invariants of complex analytic surfaces.
Since $c_1(M)=0$, $q=0$, and $p_g=1$ for a K3 surface, from (10)
we conclude that $c_2(M)=24$.  The topology of $M$ therefore
makes non-trivial topological instantons physically stable.
{}From equation (9) we see that the moduli space of
Einstein-Hermitian topological instantons, ${\cal M}_r$, on
$E\to{\rm K3}$ with topology such that $c_2(E\otimes E^*)=0$,
is of dimension zero.
\medskip
\noindent
2.  Let the flat complex two-torus, $T^2_{\bf C}={\bf C}^2/\Gamma$,
be the set of equivalence classes with respect to a given lattice
$\Gamma$.  For $T^2_{\bf C}$, the irregularity and geometric genus
take the values $q=2$, $p_g=1$, and the canonical line bundle,
$K_M$, is trivial.  From (9) it follows that the moduli space of
holomorphic connections is of complex dimension two.
Using the Noether formula (10), the topological invariants
of $T^2_{\bf C}$, and $c_1(M)=-c_1(K_M)=0$ we find that $c_2(M)=0$.
As a result, the topological instantons on the flat complex two-torus
are not topologically fixed away from the trivially flat topological
instanton---they are potentially unstable under perturbations, and
therefore are not of interest to us.
\medskip
\noindent
3. When the canonical line bundle $K_M$ is non-trivial, we turn to
the Enriques-Kodaira classification of compact, complex surfaces
(free from exceptional curves).  There are two types of compact,
complex surfaces which admit K\"ahler structures and lead to moduli
spaces of physically stable Einstein-Hermitian topological instantons
with ${\rm dim}_{\bf R}({\cal M}_r)\ge 4$ [Kod,Bar]:
\+&(a) minimal elliptic surfaces with $c_1^2(M)=0$, $c_2(M)\ge 0$, and\cr
\+&(b) minimal algebraic surfaces of `general type' with $c_1^2(M)>0$,
$c_2(M)\ge 0$.\cr
\noindent
An elliptic surface is a complex surface, $M$, with a holomorphic
projection, $\pi\colon M\to\Delta$, onto a non-singular algebraic
curve, $\Delta$, such that $\pi^{-1}(u)$ for $u\in\Delta$ is (generically)
an elliptic curve.  An algebraic surface is a compact complex surface
that can be holomorphically embedded in a projective
space, ${\bf CP}^n$.  Assume that $M$ is one or the other of
these two cases, i.e., (a) or (b).  Then the K\"ahler metric on $M$
defines a Hodge star operator, $\star$, that
splits $H^2(M,{\bf R})$ into ($\pm 1$)-eigenspaces, say
$H_\pm^2(M,{\bf R})$ of dimension $b^\pm$, where $b_2=b^++b^-$.
The signature of $M$ is defined as ${\rm sgn}(M)=b^+-b^-$.
The Hodge identities tell us that the first Betti number $b_1$ is
even, $b^+=2p_g+1$, and $h^{(1,0)}=h^{(0,1)}$ [We].
We can now rewrite equation (10) using the identity $c_1^2(M)=
3{\rm sgn}(M)+2\chi(M)$, where the Euler characteristic can be written
as $\chi(M)=2-2b_1+b_2$ (using Poincar\'e duality).
For $c_1^2(M)=0$, the Noether formula (10) is simply
$$c_2(M)= 12(p_g-q+1).\eqno{(10')}$$
If $c_1^2(M)>0$, then the Noether formula (10) becomes
$$c_2(M)= b_2+2(1-b_1).\eqno{(10'')}$$
We return to these remarks in the next section.

\bigskip
\noindent
{\bf 4. Einstein-Hermitian topological instantons on vector bundles}

\medskip
The trivial relationship between abelian topological instantons
and Einstein-Hermitian line bundles in the previous section
generalises in a not so trivial way for the vector bundle.
Using the notation in section three, $E$ is a $C^\infty$ complex
vector bundle of rank $r$ with a Hermitian structure, $h$, over
a compact complex K\"ahler surface, $(M,g,\Phi)$.  Recall from
section two that topological instantons on vector bundles with
rank greater than one are projectively flat.  Projectively
flat connections are again closely related to Einstein-Hermitian
connections, as we shall see.  If $E$ is a projectively flat
complex vector bundle of rank $r>1$, then $E$ must satisfy
$c_2(E\otimes E^*)=0$.  This leads to a simple topological
condition relating the first and second Chern classes:
$$c_2(E)={r-1\over 2r}c_1^2(E).\eqno{(11)}$$
All vector bundles in this section will be assumed to satisfy
this topological condition.  Now, every projectively flat Hermitian
vector bundle on $(M,g,\Phi)$ satisfies the weak-Einstein condition.
To see this, recall that a K\"ahler operator,
$\Lambda\colon\curv^{(p,q)}\to\curv^{(n-p,n-q)}$, can be defined [We].
The K\"ahler operator, $\Lambda$, is the adjoint operator to the
wedge multiplication of forms by the K\"ahler form, $\Phi$.  The
mean curvature, $K(D)$, can be written using the K\"ahler operator,
$$K(D)=i\Lambda H^A.$$
It is not difficult to show that all projectively flat connections
are weak-Einstein:
$$K(D)=i\Lambda H^A=i\Lambda(F\,I_E)=(i\Lambda F)I_E=\varphi I_r,$$
where we define $\varphi\equiv i\Lambda F$.
Moreover, for every projectively flat Hermitian vector bundle
$(E,h)$ there is a conformal transformation of the Hermitian
structure, $h'=ah$, for which the projectively flat Hermitian
vector bundle $(E,h')$ is Einstein-Hermitian [Ko1].
The converse is also true when (11) is satisfied.  This follows
from the L\"ubke inequality [L],
$$\int_M\,\{(r-1)c_1(E)^2-2rc_2(E)\}\wedge\Phi\le 0.\eqno{(12)}$$
The equality in (12) holds if and only if $(E,h)$ is projectively
flat [L].  Therefore the topological condition (11) implies that
Einstein-Hermitian vector bundles are projectively flat.

\smallskip
The moduli space, ${\cal M}_r={\cal E}(E,h)/U(E,h)$, of
irreducible diagonal Einstein-Hermitian $U(r)$ topological
instantons is a nonsingular K\"ahler manifold when
$${\rm deg}(M)\equiv\int_M\,c_1(M)\wedge\Phi\ge 0,$$
and $H^2(M,{\rm End}^0(E))=0$ [Kim,Ko1].  The complex
dimension of the moduli space ${\cal M}_r$, if non-empty,
depends on whether the canonical line bundle, $K_M\equiv
\Lambda^n(T^*M)$, is trivial or non-trivial.  If $K_M$
is trivial, then
$${\rm dim}_{\bf C}({\cal M}_r)=2+r^2(q-2),$$
where again $q$ is the irregularity and $r$ is the rank
of the vector bundle.  If $K_M$ is non-trivial, then
the complex dimension of the moduli space is given by
$${\rm dim}_{\bf C}({\cal M}_r)=1+r^2(q-1).$$
A useful account of Kim's work on the moduli space
of Einstein-Hermitian connections can be found in [Ko1].

\smallskip
Finally, it is interesting to note that Nahm duality
(`reciprocity' in [CG]), a property usually associated
with (anti-)instanton solutions to the (anti-)self-duality
equations, is also present for Einstein-Hermitian connections
over algebraic surfaces [Mu2,BvB,Sc].  It is this fact
that led us to the topological instanton construction in
this paper.

\bigskip
\noindent
{\bf 5. The space of motions}

\medskip
It is often stated that TFTs are physically trivial because they
have `no dynamics'.  This is due to the invariance of the
topological Lagrangian under ${\rm Diff}(M)$, so that all
point-particle world-paths can be deformed using ${\rm Diff}(M)$
to the static point-particle world-line.
This may not be true for the dynamics of Bogomol'nyi solitons,
however, because there may not exist a geometrically compatible,
static, topologically non-trivial, everywhere non-singular,
finite-Action field configuration that satisfies the Bogomol'nyi
field equations.  Put another way, the internal structure
of the soliton may provide an obstruction to trivial dynamics
where a structureless point-particle would not.  Without
explicitly solving the dynamical field equations, we can only argue
that obstructions to trivial dynamics occur by examining whether
a static `motion' is compatible with the dimension of the space
of motions.  In this section we discuss the two simplest examples
of non-trivial Einstein-Hermitian topological instanton dynamics.

\smallskip
We begin by requiring the physical stability of the instantons
under perturbation.  Assume that $M$ has a non-trivial canonical
bundle, $K_M$, so that $c_1(K_M)=-c_1(M)\ne 0$ (cf., the third
remark in section three).  If $c_1(M)>0$, then from
Kodaira's vanishing theorem and Serre duality one concludes that
$q=h^{(0,1)}=0$.  From the dimension formula (9) the
space of motions, ${\cal M}$, is either empty or of complex
dimension zero.  Therefore, in order to obtain a space of
motions of sufficient size, we require that $M$ have the
property that $c_1(M)<0$, in addition to satisfying the
inequality (8).  Now, let $M$ be a minimal elliptic ($c_1^2(M)=0$),
or, minimal algebraic ($c_1^2>0$) surface with irregularity $q=2$.
The two cases: (a) $c_1^2(M)=0$, and (b) $c_1^2(M)>0$, which
correspond to equations (10$'$) and (10$''$), respectively,
give physical stability with $q=2$, when (a) $p_g>1$, and
(b) $b_2>6$, respectively.
With $M$ taken to be space-time, the moduli space of diagonal
Einstein-Hermitian instantons, ${\cal M}_{\rm dia}$, can be
interpreted as the space of motions.
{}From the dimension formula (9) the space of motions,
when non-empty, is a real four-dimensional manifold.  Moreover,
${\cal M}_{\rm dia}$ inherits a natural holomorphic symplectic
structure from space-time $(M,g,\Phi)$ [Mu]:
$$\Theta(a,b)=\int_M\,{\rm tr}(a\wedge b)\wedge\Phi,$$
where $a,b\in T{\cal M}_{\rm dia}$.

\smallskip
Non-trival instanton solutions to the self-duality equations
in Yang-Mills theory exist only for imaginary time.
Topological instantons differ significantly from Yang-Mills
instantons since the signature of the space-time
metric has no effect on the existence of solutions.
Therefore topological instantons can have real-time dynamics.
The real dimension of the $q=2$ abelian diagonal
Einstein-Hermitian instanton space of motions (four) considered
above suggests that these topological instantons are massless free
solitons in ${\bf R}^3$ with a constant velocity [Wo]; the
dimension of the space of motions is the same dimension as the
space of (oriented) geodesics in ${\bf R}^3$.  If the velocity
were zero, the space of motions would be equivalent to the
configuration space, which would be three dimensional and not four.
Since the constant velocity of the diagonal Einstein-Hermitian
instanton appears not to vanish, the dynamics cannot be
viewed as trivial.  In effect, the topology of space-time and
the geometry of the line bundle restricts the possible field
configurations.  Diagonal Einstein-Hermitian topological
instantons might be interpreted as photons, massless neutrinos,
or stable structures made up of photons or neutrinos (ball
lightening, perhaps).

\bigskip
\noindent
{\bf 6. Conclusion}

\medskip
We have shown that there are classical generally covariant
topological field theories with a natural Bogomol'nyi structure.
The TFTs generalise Maxwell's equations and the Yang-Mills equations.
When the underlying base manifold, $M$, is taken to be a compact
K\"ahler space-time, then a natural correspondence has been
established between topological instantons and Einstein-Hermitian
connections.  The moduli space of solutions to the Bogomol'nyi
equations can be interpreted as the space of motions of the
gauge particles.  The Einstein-Hermitian condition is precisely
the condition needed to insure that this space is well-behaved.
The dimension of the space of motions depends on only the topology
of space-time and the rank of the vector bundle.  By assuming that
space-time has no exceptional curves, we have shown that space-time
is either a minimal elliptic complex surface, or, a minimal algebraic
complex surface with certain extra conditions (e.g., $c_1(M)<0$)
in order to admit Einstein-Hermitian topological solitons with
a topologically well-behaved space of motions.
We suggest that these solitons represent $n$-photon or massless
$n$-neutrino states.

\bigskip
\openup -1\jot
\centerline {{\bf References}}
\bigskip
\item{[Bar]} Barth, W., Peters, C., Van der Ven, A.: Compact
Complex Surfaces. {\it Ergebnisse der Mathematik und ihrer
Grenzgebiete: 3. Folge, Bd. 4}. Berlin: Springer-Verlag, 1984.
\smallskip
\item{[B]} Beauville, A.: Vari\'et\'es k\"ahleriennes dont la
premi\`ere classe de Chern est null. {\it J. Differ. Geom.}
{\bf 18} 755-782 (1983).
\smallskip
\item{[BS]} Baulieu, Singer: Topological Yang-Mills Symmetry. In
proceedings of conformal field theory and related topics. (Annecy,
France, March 1988).
\smallskip
\item{[BvB]} Braam, P., van Baal, P.: Nahm's transformation for
instantons. {\it Comm. Math. Phys.} {\bf 122} 267-280 (1989).
\smallskip
\item{[CG]} Corrigan, E., Goddard, P.: Construction of instanton and
monopole solutions and reciprocity, {\it Ann. Phys.} {\bf 154}
253-279 (1984).
\smallskip
\item{[D1]} Donaldson, S.K.: An application of gauge theory to the
topology of four manifolds. {\it J. Differ. Geom.} {\bf 18} 269 (1983).
\smallskip
\item{[H]} Horowitz, G.T.: Exactly soluble diffeomorphism invariant
theories. {\it Comm. Math. Phys.} {\bf 125} 417-437 (1989).
\smallskip
\item{[Kim]} Kim, H.J.: Moduli of Hermite-Einstein vector bundles.
{\it Math. Z.} {\bf 195} 143-150 (1987).
\smallskip
\item{[Ko1]} Kobayashi, S.: Differential Geometry of Complex Vector
Bundles. Princeton: Princeton U.P., 1987.
\smallskip
\item{[Ko2]} Kobayashi, S.: Curvature and stability of vector bundles.
{\it Proc. Japan Acad.} {\bf 58} 158-162 (1982).
\smallskip
\noindent
\item{[Kod]} Kodaira, K.: On the structure of compact complex analytic
surfaces. {\it Am. J. Math.} {\bf 84} 751-798 (1964).
\smallskip
\item{[L]} L\"ubke, M.: Chernklassen von
Hermite-Einstein-Vektorb\"undeln. {\it Math. Ann.} {\bf 260}
133-141 (1982).
\smallskip
\item{[Mu]} Mukai, S.: Symplectic structure of the moduli space of
sheaves on an abelian or K3 surface. {\it Invent. Math.} {\bf 77}
101-116 (1984).
\smallskip
\item{[Mu2]} Mukai, S.: Fourier functor and its application to
the moduli of bundles on a abelian variety.  Adv. Studies in Pure
Math. {\bf 10} (1987).
\smallskip
\item{[Sc]} Schenk, H.: On a generalised Fourier transform for
instantons over flat tori. {\it Comm. Math. Phys.} {\bf 116}
177-183 (1988).
\smallskip
\item{[Sou]} J-M Souriau: Structure des syst\`emes dynamique.
Paris: Dunod, 1970.
\smallskip
\item{[TRA]} Temple-Raston, M., Alexander, D.: Differential
cross-sections and escape plots for solitonic BPS $SU(2)$
magnetic monopoles. {\it Nucl. Phys.} B {\bf 397}
195-213 (1993).
\smallskip
\item{[W1]} Witten, E.: Topological Quantum Field Theory. {\it Comm.
Math. Phys.} {\bf 117} 353-386 (1988).
\smallskip
\item{[W2]} E. Witten: 2+1 dimensional gravity as an exactly soluble
system. {\it Nucl. Phys.} B {\bf 311} 46-78 (1988).
\smallskip
\item{[We]} Wells, R.O.: Differential Analysis on Complex Manifolds.
New York: Springer-Verlag, 1980.
\smallskip
\item{[Wo]} Woodhouse, N.M.J.: Geometric Quantization, 2 ed..
Oxford: Clarendon Press, 1992.
\bye